# Centralized Networked Micro Water-Energy Nexus with Proportional Exchange Among Participants

Jesus Silva-Rodriguez
Department of Electrical and Computer Engineering
University of Houston
Houston, TX, USA
jasilvarodriguez@uh.edu

Xingpeng Li
Department of Electrical and Computer Engineering
University of Houston
Houston, TX, USA
xli82@uh.edu

*Abstract*—This paper proposes a Networked Micro Water-Energy Nexus (NetMicroWEN) capable of co-optimizing and simultaneously supplying water and energy to local consumers in nearby communities. The system manages different water and energy inputs of different communities in a local network to cooperatively meet their demands. This paper considers a centralized network topology that connects all members of the network under one control system. This paper also proposes a Proportional Exchange Algorithm (PEA) that allows members to benefit equally from exchanging both resources among other members of the NetMicroWEN and the main water and power distribution systems. The co-optimization model is a mixed-integer linear program, involving all necessary power and water related constraints for the network to achieve a feasible and practical solution. The economic benefits of the NetMicroWEN are illustrated by a comparison with separate Micro Water-Energy Nexus (MWEN) systems meeting their own demands individually. The case studies demonstrate that the proposed NetMicroWEN achieves substantially lower operating costs compared to the operation of separate MWEN systems.

## Nomenclature

| | |
|---|---|
| $t$ | Time period index. |
| $T$ | Time period set. |
| $\Delta t$ | Duration of each time interval. |
| $g$ | Power generator index. |
| $G$ | Power generator set. |
| $b$ | Energy storage unit index. |
| $ES$ | Energy storage unit set. |
| $k$ | Water storage unit index. |
| $ST$ | Water storage unit set. |
| $m$ | MWEN system index. |
| $M$ | MWEN system set. |
| $P^G_{m,g,t}$ | Power output of each generator. |
| $P^G_{min,m,g}$ | Minimum power output of each generator. |
| $P^G_{max,m,g}$ | Maximum power output each generator. |
| $C^G_{m,g}$ | Generation cost of each generator. |
| $NL^G_{m,g}$ | No load cost of each generator. |
| $u^G_{m,g,t}$ | Status of each generator at every time interval |
| $SU^G_{m,g}$ | Startup cost of each generator. |
| $v^G_{m,g,t}$ | Startup indicator of each generator at every time interval. |
| $P^{grid+}_{m,t}$ | Power purchased from main grid by each MWEN. |
| $p^+_t$ | Main grid power import status at every time interval. |
| $C^{grid+}_t$ | Main grid power purchase price at every time interval. |
| $P^{grid-}_{m,t}$ | Power sold to main grid by each MWEN. |
| $p^-_t$ | Main grid power selling status at every time interval. |
| $P^{grid}_{lim}$ | Main grid-NetMicroWEN exchange power limit. |
| $C^{grid-}_t$ | Main grid power selling price at every time interval. |
| $P^{ESc}_{m,b,t}$ | Power charged to each energy storage unit. |
| $e^{ESc}_{m,b,t}$ | Charging status of each energy storage unit. |
| $P^{ESd}_{m,b,t}$ | Power discharged from each energy storage unit. |
| $e^{ESd}_{m,b,t}$ | Discharging status of each energy storage unit. |
| $\eta^{ESc}_b$ | Charging efficiency of each energy storage unit. |
| $\eta^{ESd}_b$ | Discharging efficiency of each energy storage unit. |
| $P^{ES}_{lim,m,b}$ | Charge/discharge rate limit of each energy storage unit. |
| $EL^{ES}_{m,b,t}$ | Charge level of each energy storage unit. |
| $EL^{ES}_{min,m,b}$ | Minimum charge level of each energy storage unit. |
| $EL^{ES}_{max,m,b}$ | Maximum charge level of each energy storage unit. |
| $P^N_{m,t}$ | Power exchange between each MWEN within the NetMicroWEN. |
| $P^E_{m,t}$ | Net power exchange between each MWEN and the NetMicroWEN and main grid combined. |
| $P^E_{lim,m}$ | Net power exchange limit between each MWEN and the central node of NetMicroWEN. |
| $W^{ES}_{m,b,t}$ | Water consumption rate of each energy storage unit (for hydrogen fuel cell units). |
| $P^{SP}_{m,t}$ | Solar power available at every time interval. |
| $P^{WP}_{n,t}$ | Wind power available at every time interval. |
| $P^L_{m,t}$ | Power load demand of each MWEN. |
| $W^{WW}_{m,t}$ | Water output flow rate of wastewater treatment unit. |
| $W^{WW}_{min,m}$ | Minimum water output of the wastewater treatment unit. |
| $W^{WW}_{max,m}$ | Maximum water output of the wastewater treatment unit. |
| $u^{WW}_{m,t}$ | Wastewater treatment unit status at every time interval. |
| $WL^{WW}_{m,t}$ | Wastewater reservoir level of the treatment unit. |
| $WL^{WW}_{lim,m}$ | Wastewater reservoir maximum capacity of the treatment unit. |
| $WR^{WW}_{m,t}$ | Volume flow rate of wastewater collected into reservoir of the treatment unit. |
| $NL^{WW}_m$ | Hourly operation cost of the wastewater treatment unit. |
| $CW^{WW}_m$ | Volume flow rate of wastewater treated per unit of energy. |
| $P^{WW}_{m,t}$ | Power consumption of the wastewater treatment unit. |
| $W^{WT}_{m,t}$ | Water output flow rate of regular treatment units. |
| $W^{WT}_{min,m}$ | Minimum water output of regular treatment units. |
| $W^{WT}_{max,m}$ | Maximum water output of regular treatment units. |
| $u^{WT}_{m,t}$ | Status of the regular treatment units at every time interval. |
| $NL^{WT}_m$ | Hourly operation cost of regular treatment units. |
| $CW^{WT}_m$ | Volume flow rate of water treated per unit of energy by the regular treatment units. |
| $P^{WT}_{m,t}$ | Power consumption of regular treatment units. |
| $W^{STc}_{m,k,t}$ | Water inflow of each water storage tank. |
| $W^{STd}_{m,k,t}$ | Water outflow of each water storage tank. |
| $W^{ST}_{lim,m,k}$ | Water inflow/outflow rate limit of each storage tank. |
| $sp^{ST}_{m,k,t}$ | Fill pump status of each storage tank. |
| $sv_{m,k,t}^{ST}$ | Release valve status of each storage tank. |
| $WL_{m,k,t}^{ST}$ | Water level of each storage tank at every time interval. |
| $WL_{lim,m,k}^{ST}$ | Maximum capacity of each water storage tank. |
| $W_{m,t}^{main+}$ | Water flow rate imported from main water system by each MWEN. |
| $C_t^{main+}$ | Import price of water from main water system. |
| $W_{lim}^{main}$ | Main water system-NetMicroWEN water flow rate import limit. |
| $W_{m,t}^L$ | Water load demand of each MWEN. |
| $W_{m,t}^N$ | Water exchange between each MWEN within the NetMicroWEN. |
| $W_{m,t}^E$ | Net water exchange between each MWEN and the NetMicroWEN and main water system combined. |
| $W_{lim,m}^E$ | Net water exchange limit between each MWEN and the central node of NetMicroWEN. |
| $P_{pump,m,t}^{WW}$ | Power consumption of the wastewater treatment pump. |
| $P_{pump,m,t}^{WT}$ | Power consumption of regular treatment unit pumps. |
| $P_{pump,m,k,t}^{ST}$ | Power consumption of water storage tank pumps. |

## I. INTRODUCTION

In a microgrid, various energy sources can be used to meet local power demand [1]. This allows for islanded operations, in which the microgrid can satisfy its demand when disconnected from the main grid [2-3]. Since a water network can be managed in a similar manner to a power distribution network [4], the concept of a micro water network can be explored and combined with a microgrid energy management system, forming a Micro Water-Energy Nexus (MWEN). A water-energy co-optimization model implemented for community-scale microgrids has previously been analyzed [5], in which only wastewater treatment is considered as the water input, in addition to water storage tanks and the main water system. The preliminary results in [5] demonstrate its water-energy co-optimization model achieves a lower cost of supplying both resources compared to the traditional case of a microgrid energy management system with water being supplied only by the main water network.

A MWEN co-optimizing water demand and energy consumption is proposed in [6], which also demonstrates the economic benefit of the co-optimization of both resources. However, its model only considers pump operations of a micro water distribution network as the only power consumption of the water side, without considering the consumption of local treatment processes. Another micro water-energy nexus co-optimization model is presented in [7], considering different water and energy inputs, as well as simultaneous dispatch of both resources under the same objective function. Although [7] includes water treatment from wastewater and other sources and considers their pump characteristics, it fails to consider the energy intensity of their treatment processes directly besides pump power consumption.

In [8], a renewable-based microgrid serving a drinking water treatment plant is considered, where the model optimizes water distribution along with power consumption of the treatment plant using different renewable energy inputs. However, the model does not consider power supply to end-consumers, and the only power load considered is the treatment plant. On the other hand, [9] considers the integration of different types of energy sources into a community-scale microgrid and considers their water consumption if any, but it does not consider water supply to consumers. Still, [9] explores the inclusion of hydrogen energy storage systems (HESS), further expanding upon the water-energy interdependencies in the nexus.

Another important aspect considered in this paper besides the economic benefit of multi-source water-energy co-optimization is the benefits of interconnecting multiple MWEN systems with different types of power and water sources into a network. This allows them to take advantage of a variety of sources found in different communities via an equally beneficial exchange of resources, similar to that of a network of microgrid energy management (MEM) systems. Networked microgrids can provide the additional flexibility for resilient operations by cooperatively sharing their distributed energy resources (DER) in order to overcome potential deficiencies and minimize the amount of load shed [10]. Therefore, it is reasonable to assume that a similar application in the water-energy nexus should provide the same reliability benefits. However, as seen in the literature exploring the water-energy nexus in a micro scale, there is limited studies on the possibility of further expanding the benefits of a MWEN via the implementation of a network of these systems.

To further explore and improve upon the micro water-energy nexus concept, this paper proposes a networked micro water-energy nexus (NetMicroWEN) model for a network of MWEN systems, formulated as a mixed-integer linear program (MILP). The proposed model features a Proportional Exchange Algorithm (PEA) that provides equal economic benefit to all members of the network, allowing for separate ownership of the participant MWEN. Each MWEN system will feature a variety of DERs including diesel generators, energy storage systems which can be either battery energy storage (BESS) or HESS, solar photovoltaics, wind turbines, and various distributed water resources (DWR) including water treatment units and clean water storage tanks. Each MWEN is connected with the main grid and main water network for operation during normal conditions.

The rest of this paper is organized as follows. The proposed NetMicroWEN and its optimization model are presented in Section II. Section III presents the methods employed to analyze the benefits of the proportional exchange algorithm, as well as the overall economic benefits of implementing a NetMicroWEN over the separate operation of each MWEN. Section IV presents the results of a comparison between operation of the NetMicroWEN with and without the PEA, as well as a case study exploring the economic comparison between the use of a network over separate systems. Finally, section V concludes the paper.

## II. MODEL DESCRIPTION

The idea of a NetMicroWEN is to be able to interconnect different individual nearby local systems that own different resources to collaboratively meet their combined electricity and water demands. A diagram of the NetMicroWEN based on a central network topology displaying the power (red) and water (blue) flows among MWEN and the main grid and water system is shown in Fig. 1.

In a central network topology for microgrids, all resources are scheduled by a central energy management system for optimal energy sharing between the microgrids in the network

[11]. The centralized NetMicroWEN follows that same principle, and a central water-energy management system carries out the optimization with information from all the MWEN in the network, achieving a global objective of minimizing the combined operating costs of all MWEN together. This objective function is formulated as (1), where the direct operation cost associated with power and water inputs are calculated by (2) and (3) respectively.

$$\text{minimize } \sum_{m \in M} f_{cost,m} = \sum_{m \in M} [f_{E,m} + f_{W,m}] \quad (1)$$

$$f_{E,m} = \sum_{t \in T} \{ \sum_{g \in G} [SU_{m,g}^G v_{m,g,t}^G + \Delta t \cdot (NL_{m,g}^G u_{m,g,t}^G + C_{m,g}^G P_{m,g,t}^G)] + C_t^{grid+} P_{m,t}^{grid+} - C_t^{grid-} P_{m,t}^{grid-} + C_t^{Np} L_{m,t}^N \} \quad (2)$$

$$f_{W,m} = \sum_{t \in T} \{ \Delta t \cdot (NL_m^{WW} u_{m,t}^{WW} + NL_m^{WT} u_{m,t}^{WT}) + C_t^{main+} W_{m,t}^{main+} + C_t^{Nw} D_{m,t}^N \} \quad (3)$$

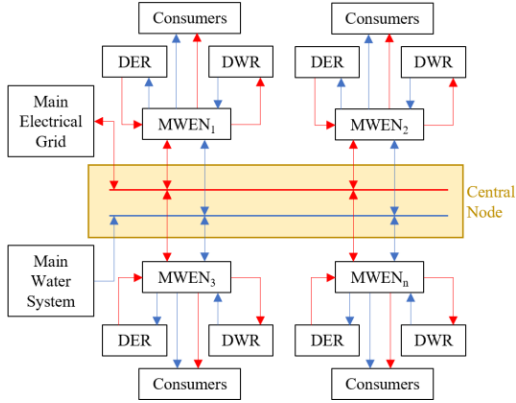

Fig. 1: NetMicroWEN in a central network topology configuration.

The system constraints are defined mostly based on local characteristics of the sources and components of each MWEN, as well as some global system constraints that regulate the exchanges within the network and the main distribution systems. These constraints can be grouped into power related constraints (4)-(19) and water related constraints (20)-(35). Constraint (4) regulates the output power of the diesel generators while (5) keeps track of the status of each generator. Constraints (6)-(7) limit the charging and discharging rate of the energy storage systems, and (8) ensures energy storage units are only either charging or discharging or staying idle at each time interval; (9) determines the energy level of each storage unit at each time interval, and (10) limits the energy level of each unit. Constraint (11) is for the HESS units, and it determines the water consumption rate of each HESS depending on the power at which it is being charged, representing the water needed to perform electrolysis and store energy in the form of hydrogen. Constraint (12) sets the power balance of each MWEN, and (13) simply determines the net power load of each MWEN based on the power consumed by the water components as well. Constraints (14)-(15) regulate the net power $P_{m,t}^E$ flowing in/out of each MWEN, where $P_{m,t}^E$ is a combination of the power exchanged with other MWENs and the power exchanged with the main grid, since they both flow through the central node. Constraint (16) is a global constraint that establishes a power balance of the network power exchanges among participants. Constraints (17)-(18) regulate the direct import and export of power between the NetMicroWEN and the main grid, and (19) ensures power cannot be imported and exported simultaneously. Note that (17)-(19) are binding only when the main grid tie-line limit is lower than the sum of the tie-line limits of each MWEN and the central node, otherwise the net main grid import or export would be limited by the sum of each MWEN tie-line rather than its own tie-line limit.

$$P_{min,m,g}^G u_{m,g,t}^G \leq P_{m,g,t}^G \leq P_{max,m,g}^G u_{m,g,t}^G, \forall m \in M, g \in G, t \in T \quad (4)$$

$$v_{m,g,t}^G \geq u_{m,g,t}^G - u_{m,g,t-1}^G, \forall m \in M, g \in G, t \in T \quad (5)$$

$$0 \leq P_{m,b,t}^{ESc} \leq P_{lim,m,b}^{ES} e_{m,b,t}^{ESc}, \forall m \in M, b \in ES, t \in T \quad (6)$$

$$0 \leq P_{m,b,t}^{ESd} \leq P_{lim,m,b}^{ES} e_{m,b,t}^{ESd}, \forall m \in M, b \in ES, t \in T \quad (7)$$

$$e_{m,b,t}^{ESc} + e_{m,b,t}^{ESd} \leq 1, \forall m \in M, b \in ES, t \in T \quad (8)$$

$$EL_{m,b,t}^{ES} = EL_{m,b,t-1}^{ES} + \Delta t \cdot \left( \eta_{m,b}^{ESc} P_{m,b,t}^{ESc} - P_{m,b,t}^{ESd} / \eta_{m,b}^{ESd} \right), \forall m \in M, b \in ES, t \in T \quad (9)$$

$$EL_{min,m,b}^{ES} \leq EL_{m,b,t}^{ES} \leq EL_{max,m,b}^{ES}, \forall m \in M, b \in ES, t \in T \quad (10)$$

$$W_{m,b,t}^{ES} = CW_{m,b}^{ES} P_{m,b,t}^{ESc}, \forall m \in M, b \in ES, t \in T \quad (11)$$

$$\sum_{g \in G} P_{m,g,t}^G + \sum_{b \in ES} [P_{m,b,t}^{ESd} - P_{m,b,t}^{ESc}] + P_{m,t}^{grid+} - P_{m,t}^{grid-} + P_{m,t}^N = P_{m,t}^{net}, \forall m \in M, t \in T \quad (12)$$

$$P_{m,t}^{net} = P_{m,t}^L - P_{m,t}^{SP} - P_{m,t}^{WP} + P_{m,t}^{WD} + P_{pump,m,t}^{WD} + P_{m,t}^{WT} + P_{pump,m,t}^{WT} + \sum_{k \in ST} P_{pump,m,k,t}^{ST}, \forall m \in M, t \in T \quad (13)$$

$$P_{m,t}^E = P_{m,t}^{grid+} - P_{m,t}^{grid-} + P_{m,t}^N, \forall m \in M, t \in T \quad (14)$$

$$-P_{lim,m}^E \leq P_{m,t}^E \leq P_{lim,m}^E, \forall m \in M, t \in T \quad (15)$$

$$\sum_{m \in M} [P_{m,t}^N] = 0, \forall t \in T \quad (16)$$

$$0 \leq \sum_{m \in M} P_{m,t}^{grid+} \leq P_{lim}^{grid} p_t^+, \forall t \in T \quad (17)$$

$$0 \leq \sum_{m \in M} P_{m,t}^{grid-} \leq P_{lim}^{grid} p_t^-, \forall t \in T \quad (18)$$

$$p_t^+ + p_t^- \leq 1, \forall t \in T \quad (19)$$

For the water-related constraints, (20) limits the output flow rate of the wastewater treatment units, while (21) determines the water level of the untreated wastewater reservoir at every time interval with the recovery rate set to a percentage of the water demand of the previous time interval; (22) limits the capacity of the reservoir, and (23) sets the relation between the output flow rate and the power consumption of the wastewater treatment unit. Constraints (24)-(25) set the output flow rate limit and power consumption relation for freshwater treatment and/or water desalination units, depending on which are available in each MWEN. Constraints (26)-(27) limit the water flow of the water storage pump and release valve, respectively, and (28) ensures the tanks are only either being filled or water is being drawn from them at every time interval. Constraint (29) determines the water level of each storage tank at every time interval, while (30) limits their capacity. Constraint (31) is the water balance constraint, which ensures total water output matches the net water demand for each MWEN, and (32)-(33) limit the net water flow between each MWEN and the central node of the NetMicroWEN, for local water network exchanges as well as main water system imports. Constraint (34) establishes the global water balance of the network, and (35) sets the limit of the coupling between the central node and the main water system. A bidirectional exchange between the NetMicroWEN and the main water system is not considered because previous preliminary results have shown that, because water prices are constant throughout the day, there is no benefit of importing water at low peak hours and exporting at high peak hours, which is different from the practices of power exchange between microgrids and the main grid [5].

$$W_{min,m}^{WW} u_{m,t}^{WW} \leq W_{m,t}^{WW} \leq W_{max,m}^{WW} u_{m,t}^{WW}, \forall m \in M, t \in T \quad (20)$$

$$WL_{m,t}^{WW} = WL_{m,t-1}^{WW} + \Delta t \cdot \left(WR_{m,t}^{WW} - W_{m,t}^{WW}\right), \forall m \in M, t \in T \quad (21)$$

$$0 \leq WL_{m,t}^{WW} \leq WL_{lim,m}^{WW}, \forall m \in M, t \in T \quad (22)$$

$$W_{m,t}^{WW} = CW_m^{WW} P_{m,t}^{WW}, \forall m \in M, t \in T \quad (23)$$

$$W_{min,m}^{WT} u_{m,t}^{WT} \leq W_{m,t}^{WT} \leq W_{max,m}^{WT} u_{m,t}^{WT}, \forall m \in M, t \in T \quad (24)$$

$$W_{m,t}^{WT} = CW_m^{WT} P_{m,t}^{WT}, \forall m \in M, t \in T \quad (25)$$

$$0 \leq W_{m,k,t}^{STc} \leq W_{lim,m,k}^{ST} sp_{m,k,t}^{ST}, \forall m \in M, k \in ST, t \in T \quad (26)$$

$$0 \leq W_{m,k,t}^{STd} \leq W_{lim,m,k}^{ST} sv_{m,k,t}^{ST}, \forall m \in M, k \in ST, t \in T \quad (27)$$

$$sp_{m,k,t}^{ST} + sv_{m,k,t}^{ST} \leq 1, \forall m \in M, k \in ST, t \in T \quad (28)$$

$$WL_{m,k,t}^{ST} = WL_{m,k,t-1}^{ST} + \Delta t \cdot \left(W_{m,k,t}^{STc} - W_{m,k,t}^{STd}\right), \forall m \in M, k \in ST, t \in T \quad (29)$$

$$0 \leq WL_{m,k,t}^{ST} \leq WL_{lim,m,k}^{ST}, \forall m \in M, k \in ST, t \in T \quad (30)$$

$$W_{m,t}^{WW} + W_{m,t}^{WT} + \sum_{k \in ST}\left[W_{m,k,t}^{STd} - W_{m,k,t}^{STc}\right] + W_{m,t}^{main+} + W_{m,t}^{N} = W_{m,t}^{L} + \sum_{b \in ES} W_{m,b,t}^{ES}, \forall m \in M, t \in T \quad (31)$$

$$W_{m,t}^{E} = W_{m,t}^{main+} + W_{m,t}^{N}, \forall m \in M, t \in T \quad (32)$$

$$-W_{lim,m}^{E} \leq W_{m,t}^{E} \leq W_{lim,m}^{E}, \forall m \in M, t \in T \quad (33)$$

$$\sum_{m \in M}\left[W_{m,t}^{N}\right] = 0, \forall t \in T \quad (34)$$

$$0 \leq \sum_{m \in M} W_{m,t}^{main+} \leq W_{lim}^{main}, \forall t \in T \quad (35)$$

Additionally, as part of the water-related constraints, there are also a group of equality constraints that are defined to establish the relation between water flow rate and the power consumption of the water pumps of all treatment units and water storage systems. These are calculated using (36) which is a linearized expression determined based on relations between hydraulic head gain of water pumps and their produced water flow rate [7], as well as pump curves for a common water pump [12]. In (36), $\eta$ represents the pump's efficiency, and $\alpha$ represents the equivalent linear relation between flow rate and power consumption.

$$\eta P_{pump} = \alpha \cdot W \quad (36)$$

## III. METHODOLOGY

The NetMicroWEN proposed in this paper is meant to accomplish two objectives. The first objective is to establish a centralized water-energy management network of MWENs that can achieve a fair and equally beneficial participation for all systems in the network. The second objective is to yield overall economic benefits for the MWENs participating in the network, compared to the results if they were to operate as separate systems without any coordination.

### A. Proportional Exchange Algorithm

When the NetMicroWEN optimization model presented in section II is executed, there may exist multiple optimal solutions, however, not all of them will result in the lowest individual operating cost for each MWEN. In order to adjust the solution to one that preserves the same overall network operating cost but fairly balances each individual MWEN cost, this paper proposes the PEA algorithm presented in Algorithm 1. The cost of buying or selling power and water between the MWEN in the network is set to a price in between the purchasing and selling prices of the main distribution systems in order to encourage collaboration within the network. Therefore, the algorithm adjusts the current solution from the optimization model by checking how much total power or water each MWEN requires to import or export, and then balances these needs from what is available within the network and meets any remaining needs assigning an import/export with the main systems.

| Algorithm 1: Proportional Exchange Algorithm for power exchange |
|---|
| 1. Solve the NetMicroWEN MILP problem and obtain network exchanges $L_{m,t}^{N}$ and net MWEN exchanges $L_{m,t}^{E}$. |
| 2. **For** $t$ in $T$ |
| 3.    **For** $m$ in $M$ |
| 4.       **If** $L_{m,t}^{E} > 0$ |
| 5.          Set $L_{m,t}^{E+} = L_{m,t}^{E}$ and $L_{m,t}^{E-} = 0$ |
| 6.       **Else** Set $L_{m,t}^{E+} = 0$ and $L_{m,t}^{E-} = L_{m,t}^{E}$ |
| 7.    **end For** |
| 8.    **For** $m$ in $M$ |
| 9.       **If** $L_{m,t}^{E} > 0$ |
| 10.         **If** $\left|\sum_{m \in M} L_{m,t}^{E+}\right| > \left|\sum_{m \in M} L_{m,t}^{E-}\right|$ |
| 11.            Set $L_{m,t}^{N} = \frac{L_{m,t}^{E+}}{\left|\sum_{m \in M} L_{m,t}^{E+}\right|} \cdot \left|\sum_{m \in M} L_{m,t}^{E-}\right|$ |
| 12.         **Else** Set $L_{m,t}^{N} = L_{m,t}^{E}$ |
| 13.         Set $P_{m,t}^{grid+} = L_{m,t}^{E} - L_{m,t}^{N}$ and $P_{m,t}^{grid-} = 0$ |
| 14.         **Else If** $\left|\sum_{m \in M} L_{m,t}^{E+}\right| < \left|\sum_{m \in M} L_{m,t}^{E-}\right|$ |
| 15.            Set $L_{m,t}^{N} = \frac{L_{m,t}^{E-}}{\left|\sum_{m \in M} L_{m,t}^{E-}\right|} \cdot \left|\sum_{m \in M} L_{m,t}^{E+}\right|$ |
| 16.         **Else** $L_{m,t}^{N} = L_{m,t}^{E}$ |
| 17.         Set $P_{m,t}^{grid-} = \left|L_{m,t}^{E} - L_{m,t}^{N}\right|$ and $P_{m,t}^{grid+} = 0$ |
| 18.    **end For** |
| 19. **end For** |

While Algorithm 1 shows PEA for power exchange only, it can effectively be applied for water exchange as well using the equivalent water variables.

### B. Economic Analysis

To test and verify the economic improvements of the NetMicroWEN over separate operations of MWEN systems without coordination, a 24-hour time period $T$ is considered. The optimization problem for the NetMicroWEN is performed in normal operation with the central node connected to the main power grid and water system. The final values of water and energy storage levels are set to a percentage of their respective peak demands of that time period for reliability purposes. Moreover, the final wastewater reservoir level will also be restricted to be less than or equal to a percentage of its maximum capacity in order to require systems to treat wastewater as an ecological requirement.

The same procedure is performed for each MWEN individually without cooperation among each other. Their individual operating costs as well as the total cost over all MWENs are compared with those of the NetMicroWEN case.

## IV. CASE STUDIES

To explore the economic benefits of the NetMicroWEN as well as the benefits of the PEA for every MWEN in the network, a test case composed of four MWEN systems following the structure of Fig. 1 is developed. Each system has a different variety of DER and DWR, as well as different water [13] and power [14] demand profiles and RES power available [15-16], which will be included in the net power and water loads of each of the analyses of this section. Table I shows parameters for the DERs and Table II is for the DWRs of each MWEN of this test case.

Table I: Test case DER parameters.

| DER Property | MWEN$_1$ | MWEN$_2$ | MWEN$_3$ | MWEN$_4$ |
|---|---|---|---|---|
| $\sum_{g \in G} P^G_{min,m,g}$ | 1450 [kW] | 2390 [kW] | 900 [kW] | 0 |
| $Avg_{g \in G}(C^G_{m,g})$ [17] | 0.305 [$/kWh] | 0.28 [$/kWh] | 0.26 [$/kWh] | 0 |
| $Avg_{g \in G}(NL^G_{m,g})$ | 9.85 [$/h] | 9.03 [$/h] | 8.45 [$/h] | 0 |
| $Avg_{g \in G}(SU^G_{m,g})$ | 14.00 [$] | 12.78 [$] | 11.85 [$] | 0 |
| $\sum_{b \in ES} P^{ES}_{lim,m,b}$ | 1800 [kW] | 2900 [kW] | 1500 [kW] | 3625 [kW] |
| $\sum_{b \in ES} EL^{ES}_{max,m,b}$ | 9960 [kWh] | 10000 [kW] | 8300 [kWh] | 12500 [kW] |
| $\eta^{ESc}$ | 80% | 95% | 80% | 95% |
| $\eta^{ESd}$ | 60% | 98% | 60% | 98% |

Table II: Test case DWR parameters.

| DWR Property | MWEN$_1$ | MWEN$_2$ | MWEN$_3$ | MWEN$_4$ |
|---|---|---|---|---|
| $W^{WW}_{max,m}$ | 720 [gal/h] | 1000 [gal/h] | 720 [gal/h] | 1000 [gal/h] |
| $WL^{WW}_{lim,m}$ | 12000 [gal] | 18600 [gal] | 12000 [gal] | 18600 [gal] |
| $CW^{WW}_m$ [18] | 365 [gal/kWh] | 382 [gal/kWh] | 365 [gal/kWh] | 382 [gal/h] |
| $NL^{WW}_m$ | 75 [$/h] | 75 [$/h] | 75 [$/h] | 75 [$/h] |
| $W^{WT}_{max,m}$ | 1500 [gal/h] | 525 [gal/h] | 1730 [gal/h] | 0 |
| $CW^{WT}_m$ [18] | 96 [gal/kWh] | 4400 [gal/kWh] | 100 [gal/kWh] | 0 |
| $NL^{WT}_m$ | 17.20 [$/h] | 25.50 [$/h] | 15.00 [$/h] | 0 |
| $\sum_{k \in ST} W^{ST}_{lim,m,k}$ | 900 [gal/h] | 900 [gal/h] | 900 [gal/h] | 900 [gal/h] |
| $\sum_{k \in ST} WL^{ST}_{lim,m,k}$ | 10209 [gal] | 10209 [gal] | 10209 [gal] | 10209 [gal] |

Each MWEN will have tie-line limits of 1400 kW and 980 gal/h for power and water, respectively, and the wastewater recovery rate is set to 50%. The tie-line limits between the central node and the main power and water systems are set to the sum of the individual MWEN tie-lines, that is 5600 kW and 3920 gal/h, respectively, for the NetMicroWEN case. This is done so in order to allow each system to have access to the same amount of electric power and water at every time interval on both cases. Moreover, it is important to note that MWEN 1 and 3 are considered to be coastal communities, and for this reason these MWENs feature offshore wind in addition to some solar power, implement HESS, and use water desalination units, while MWEN 2 and 4 implement BESS and only solar as their RES. Also, MWEN 2 features freshwater treatment. The energy storage efficiencies of MWEN 1 and 3 are much lower because they represent those of the HESS electrolysis [19] and fuel cell [20]. Also, the hourly electricity price is based on ERCOT's locational marginal price database [21], and the cost of water is set to a constant $0.006/gal based on the Texas Municipal League information [22].

### A. Proportional Exchange Algorithm

The NetMicroWEN optimization model of section II is executed without the PEA, and then is executed again but now implementing the PEA. The power exchanges of each MWEN within the network are presented in Fig. 2 and 3, respectively.

These results show how the exchanges within the network without the PEA seem to be mostly dominated by one of the MWEN every time, causing one MWEN to enjoy the economic benefits of the network more than the other systems. However, with the implementation of the PEA, it can be noted how for example the power exported by MWEN 4, it is now imported by other systems in a more proportional way in the later hours of the day. A similar outcome is seen with water exchange within the network; power and water exchanges with the main systems also follow this pattern, in which each MWEN receives a more balanced trade of resources to produce a fair exchange opportunity.

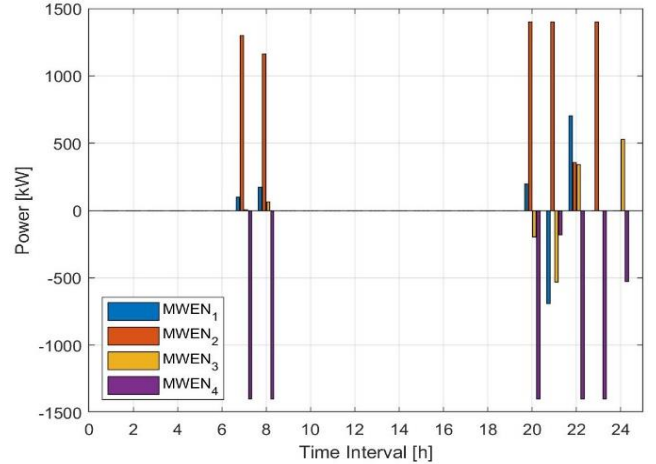

Fig. 2: Network power exchanges for case without PEA.

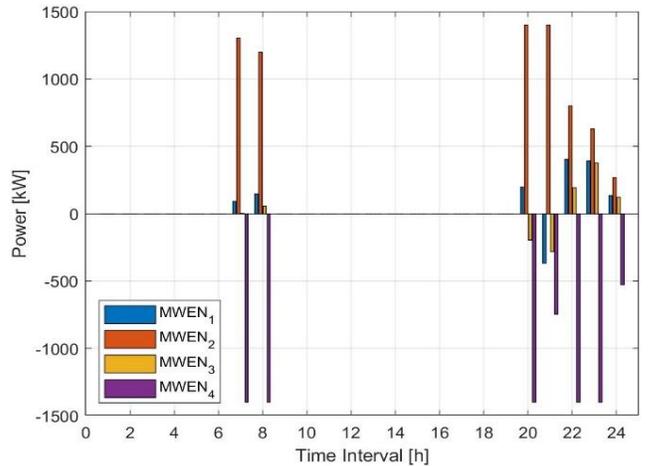

Fig. 3: Network power exchanges for case with PEA.

These variations in the exchanges of all MWEN yield a more balanced economic benefit for all participants in the network, and Table III shows the difference in operating costs of each individual MWEN during the case without PEA and the case with PEA, to show how these were balanced after the algorithm implementation.

Table III: MWEN individual operating costs with and without PEA

| MWEN | Cost before PEA | Cost after PEA | % Difference |
|---|---|---|---|
| 1 | $1648.70 | $1654.94 | -0.378% |
| 2 | $3392.42 | $3384.39 | 0.237% |
| 3 | $1519.05 | $1529.04 | -0.658% |
| 4 | $417.32 | $409.12 | 1.965% |
| TOTAL | $6977.49 | 6977.49 | 0% |

## B. Economic Comparison

The NetMicroWEN optimization is set for comparison with the optimization of separate MWEN systems. Both cases are assumed to have the same load and RES profiles. Table IV shows the results for the individual operating costs for the NetMicroWEN case and the case for each MWEN operating separately, as well as the total cost reduction of all four MWEN combined. The results show a substantial reduction in operating costs for each MWEN, proving all systems benefit by participating in a NetMicroWEN. In both cases, each system had the same access to the main grid and water systems, which are the sources with the lowest cost, however, gaining access to other MWENs resources provides for a greater economic advantage and more flexible water-energy management strategies, resulting in substantial operating cost reductions for each MWEN individually, as well as for all four systems combined.

Table IV: Individual MWEN operating costs for the NetMicroWEN and separate MWEN operation cases.

| MWEN | Separate MWEN cost | NetMicroWEN cost | % Difference |
|---|---|---|---|
| 1 | $1916.54 | $1603.14 | 16.35% |
| 2 | $3764.35 | $3371.93 | 10.42% |
| 3 | $1759.67 | $1439.91 | 18.17% |
| 4 | $656.10 | $367.60 | 43.97% |
| TOTAL | $8096.66 | $6782.58 | 16.23% |

## V. CONCLUSIONS

The proposed Networked Micro Water-Energy Nexus is capable of co-optimizing water and energy by managing the resources of multiple communities participating in the network. The NetMicroWEN optimization model by itself minimizes the combined cost of all MWENs in the network, however it does not provide the fairest economic benefit to each participant. To resolve this issue, a Proportional Exchange Algorithm is proposed, balancing the individual costs of each MWEN while preserving the optimal value of the combined cost across all MWENs. The NetMicroWEN was also compared to the same number of MWENs operating separately with no coordination. The comparison was carried out for a 24-hour period in normal operation which allowed an economic comparison. The results proved that the NetMicroWEN can obtain individual MWEN operating cost reductions, with a net reduction of 16.23% across all systems, compared to each system operating separately. In summary, the NetMicroWEN proposed in this paper is able to effectively supply both electricity and water demands of nearby local communities while providing fair individual cost benefits to each participant, in addition to a collective reduction in costs for all systems participating in the network.